\documentclass[fleqn,twoside]{article}
\usepackage{espcrc2}

\usepackage{epsfig}

\newcommand{\AmS}{{\protect\the\textfont2
  A\kern-.1667em\lower.5ex\hbox{M}\kern-.125emS}}

\title{SU(2) Lattice Gauge Theory at Nonzero Chemical Potential and Temperature.}

\author{J.~B.~Kogut\address{Physics Department, University of Illinois,
                            1110 West Green Street, Urbana, IL 61801, USA},
        D.~Toublan\addressmark
   and  D.~K.~Sinclair\address{HEP Division, Argonne National Laboratory,
                               9700 South Cass Avenue, Argonne, IL 60439, USA}
\thanks{Talk presented by JBK.
JBK and DT are supported in part by an NSF grant NSF PHY-0102409.
D.T. is supported in part by ``Holderbank''-Stiftung.
DKS is supported by DOE contract
W-31-109-ENG-38.
Simulations used IBM SPs and CRAYs at NERSC and NPACI.}}

\begin{document}
\begin{abstract}
$SU(2)$ lattice gauge theory with four flavors of quarks is simulated at nonzero
chemical potential $\mu$ and temperature $T$ and the results are
compared to the predictions of Effective Lagrangians. Simulations
on $16^4$ lattices indicate that at zero $T$ the theory experiences a second
order phase transition to a diquark condensate state which is well described by mean field theory.
Nonzero $T$ and $\mu$ are studied on $12^3 \times 6$ lattices. For
low $T$, increasing $\mu$ takes the system through a line of second order
phase transitions to a diquark condensed phase. Increasing $T$ at high $\mu$,
the system passes through a line of first order transitions from the diquark
phase to the quark-gluon plasma phase.
\end{abstract}

\maketitle

\section{Introduction}

Since $SU(3)$ QCD cannot be simulated or studied analytically at moderate
chemical potentials,
theorists have turned to simpler models. One of the more interesting is the
color $SU(2)$ version of QCD \cite{Shuryak}, \cite{Wilczek}, \cite{Hands}.
In this model diquarks do not
carry color, so their condensation does not break color symmetry dynamically.
The critical chemical potential is
one-half the mass of the lightest meson, the pion, because quarks and
anti-quarks reside in equivalent representations of the $SU(2)$ color group.
Chiral Lagrangians can be used to study the diquark condensation transition in
this model because the critical chemical potential vanishes in the chiral
limit, and the model has a Goldstone realization of the spontaneously broken
quark-number symmetry \cite{Toublan,SUNY,STV1,SSS}. The problem has
also been 
studied within a Random Matrix Model at non-zero $\mu$ and $T$
\cite{Vanderheyden}. Lattice simulations of the model are also possible
because the fermion determinant is real and non-negative for all chemical
potentials.

Preliminary lattice simulations of the $SU(2)$ model with four species of
quarks, simulation data and an Effective Lagrangian analysis of aspects of
the $T$-$\mu$ phase diagram were recently published \cite{DK2000,PLB}.
Very early work on this model at
finite $T$ and $\mu$ was performed by \cite{t+mu}. A simulation study of the
spectroscopy of the light bosonic modes will be presented elsewhere
\cite{inprep}. This work is based on \cite{KTD2}.

In our exploratory study \cite{PLB}, we found
a line of transitions surrounding a phase with a diquark condensate. Along this line
there is a tricritical point, labeled $2$ in Fig. 1, where the transition
switches from being second order to first order. We will present evidence for metastability
along the line at high $\mu$.
The tricritical point $2$, has a natural explanation in
the context of chiral Lagrangians \cite{PLB}. Following the formalism of \cite{Toublan}
we argued that trilinear couplings among the low lying boson fields of the
Lagrangian become more significant as $\mu$ and $T$ increase and they can
cause the transition to become first order at a $\mu$ value in the vicinity of
the results found in the simulation.
A tricritical point is also found in Chiral Perturbation
Theory \cite{STV2}.


\begin{figure}[htb!]
\centerline{
\epsfxsize 3 in
\epsfysize 3 in
\epsfbox{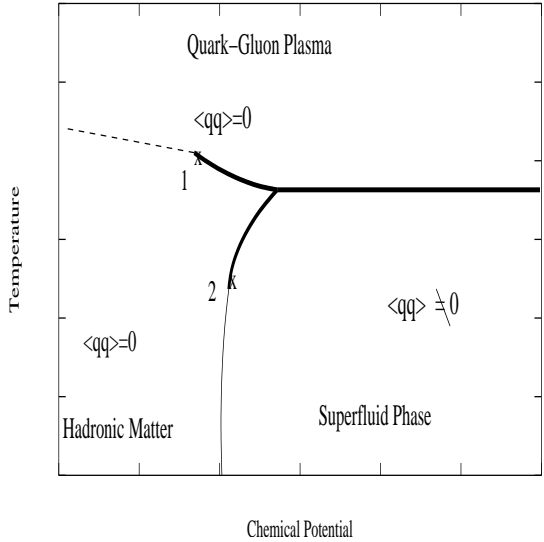}
}
\caption{Schematic Phase Diagram in the $T$-$\mu$ Plane. The thin(thick) line
consists of second(first) order transitions. The dashed line denotes a crossover.
Point 1 labels a critical point and Point 2 labels a tricritical point.
The existence and position of point 1 is
a matter of conjecture in the two color model.
}
\label{fig:phase}
\end{figure}


The conjectured phase diagram of Fig.~\ref{fig:phase} also has a critical
point labeled $1$ which connects the line of first order transitions to the
dashed line of crossover phenomena extending to the $\mu=0$ axis where we
expect a finite T chiral crossover between conventional hadronic matter and
the quark-gluon plasma. For sufficiently light quarks in the four flavor SU(2)
model, the $\mu=0$ transition is known to be first order both from
theoretical arguments \cite{PW,Wirstam} and simulations \cite{JBK}. For quark
mass $m=0.05$ we find that the transition is smoothed out to a crossover.
If the critical point 1 would be absent, a not unlikely possibility, the crossover line would
intersect the curve separating the diquark condensation phase from
the normal phase.

\section{Simulation Results and Analysis}

Consider the simulation results for the $N_f=4$ theory on $16^4$
lattices. We simulated the $SU(2)$ model at a 
relatively weak coupling $\beta=1.85$, within
the theory's apparent scaling window, in order to make contact with
the theory's continuum limit. The quark mass was $m= 0.05$, as in \cite{PLB},
and a series of simulations were performed at diquark source strength 
$\lambda = 0.0025$, $0.005$, and
$0.01$ so that our results could be extrapolated to vanishing diquark source,
$\lambda=0$.

\begin{figure}[htb!]
\centerline{
\epsfxsize 3 in
\epsfysize 3 in
\epsfbox{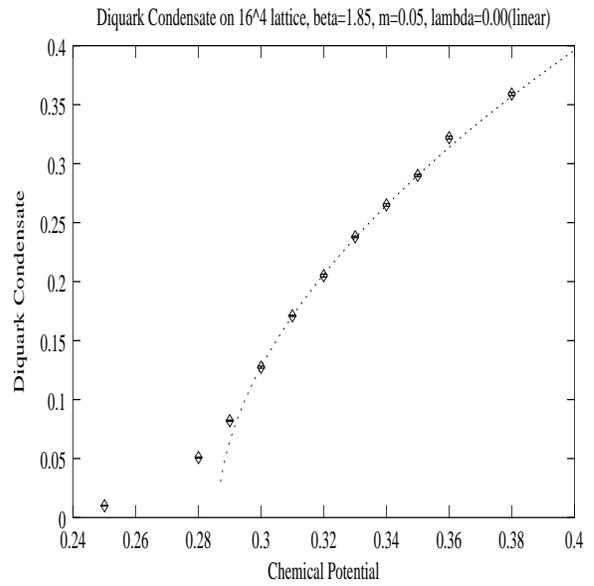}
}
\caption{Diquark Condensate vs. $\mu$}
\label{fig:linear}
\end{figure}

In Fig.~\ref{fig:linear} we show the diquark condensate ($\langle\chi^T\tau_2\chi\rangle$), 
linearly extrapolated to $\lambda=0$, plotted against the
chemical potential $\mu$.

\noindent We see evidence to a quark-number violating second order phase
transition in this figure. The dashed line is a power law fit from $\mu=0.30$
to $0.35$ which predicts the critical chemical potential of $\mu_c =
0.2860(2)$. The power law fit is good, its confidence level is $48$ percent,
and its critical index is $\beta_{mag}= 0.54(3)$ which is consistent
with the mean field result $\beta_{mag}=1/2$, predicted by chiral perturbation
theory \cite{Toublan}, \cite {SUNY} including one loop corrections as well
as simulations of 2-color QCD in the strong coupling
limit \cite{aadgg}.

\section{Runs at finite temperature}

Now consider $12^3 \times 6$ simulations.
Past $8^3 \times 4$ simulations and Effective Lagrangian
analyses predicted that there is a line of first order
transitions at high $\mu$ and high $T$ \cite{PLB}.

\begin{figure}[htb!]
\centerline{
\epsfxsize 3 in
\epsfysize 3 in
\epsfbox{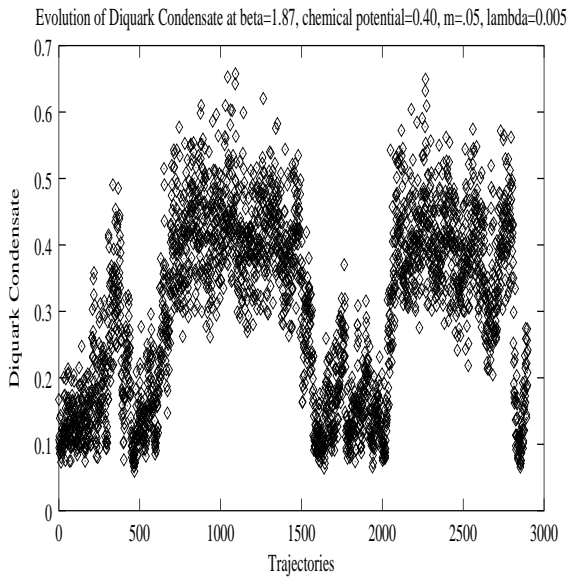}
}
\caption[]{Diquark Condensate vs. Computer Time.}
\label{fig:meta}
\end{figure}

Our best evidence for a first order transition comes from the
time evolution of the observables
at $\beta=1.87$ and $\mu \geq 0.40$ which show signs of metastability. For example, in the
figure~\ref{fig:meta} we show the time evolution
of the diquark condensate and display several tunnelings between a
state having a condensate
near $0.15$ and another with a condensate near $0.40$.

\section{Conclusions}

We are currently calculating the model's pseudo-Goldstone boson mass spectrum
and instanton content. We hope that these measurements will add more insight into
this model's curious phase transitions.

\end{document}